\documentclass{moriond}

\usepackage{amsmath, cite, hyperref, listings}
\usepackage[dvipsnames]{xcolor}
\hypersetup{colorlinks=true, linkcolor=Maroon, citecolor=BurntOrange, filecolor=BurntOrange, urlcolor=BurntOrange}
\def\be{\begin{equation}}
\def\ee{\end{equation}}
\def\bea{\begin{eqnarray}}
\def\eea{\end{eqnarray}}

\newcommand{\Photo}{}

\begin{document}
\vspace*{4cm}
\title{Toponium physics at the Large Hadron Collider}

\author{Benjamin Fuks}

\address{\vspace{.2cm}
  $^1$ Laboratoire de Physique Théorique et Hautes Énergies (LPTHE), UMR 7589, Sorbonne Université et CNRS, 4 place Jussieu, 75252 Paris Cedex 05, France}

\maketitle\abstracts{We examine toponium formation effects in top-antitop pair production at the LHC, focusing on the near-threshold region where non-relativistic corrections are relevant. We discuss their modelling using non-relativistic QCD Green's functions, and show that predictions reproduce features expected from bound-state dynamics, in contrast to pseudo-scalar toy models.}

\section{Introduction}
The formation of toponium, a colour-singlet bound state of a top and an antitop quark, was predicted well before the discovery of the top quark~\cite{Fadin:1987wz}. In this case, due to the large top mass $m_t$, toponium states decay almost instantaneously and do not produce the sharp resonance peaks observed in other quarkonium systems. As a result, their impact on top-antitop production has long been considered experimentally unobservable. However, recent excesses in several distributions of top-antitop events in LHC data~\cite{ATLAS:2023gsl, CMS:2024ybg, CMS:2025kzt} have been found to be consistent with the expectations for toponium formation~\cite{Fuks:2021xje}. To investigate this possibility, toponium effects must be computed through the Green's function of the non-relativistic QCD Hamiltonian~\cite{Fadin:1990wx, Hagiwara:2008df}, although the leading contributions can in principle be approximated using a toy model in which toponium is represented by a pseudo-scalar resonance coupling to gluons and top quarks. In this report, we compare two toponium modelling strategies, a first one based on introducing a fictitious resonance representing the toponium states~\cite{Fuks:2021xje, Maltoni:2024tul}, and another based on re-weighting the relevant matrix elements using Green's functions derived from non-relativistic QCD~\cite{Fuks:2024yjj}.

\section{Toponium formation at hadronic colliders}\label{sec:topoform}
We consider the production of a top-antitop pair at a space-time point $x$, and top and antitop decays at positions $y$ and $z$ respectively. The corresponding three-point Green's function $K(x,y,z)$ can be written as~\cite{Sumino:1992ai}
\begin{equation}
\begin{split}
  K_{abcd}(x,y,z) =&\ \frac{(1+\gamma^0)_{ca}}{2}\, \frac{(1-\gamma^0)_{bd}}{2} \int \mathrm{d}^3r \Big[ K_1\big(y; (z^0, \vec{r})\big)\,  K_2(z^0, \vec{r}, \vec{z}; x^0, \vec{x}, \vec{x}) \\
  &\qquad\qquad\qquad + K_1\big(z; (y^0, \vec{r})\big)\, K_2(y^0, \vec{y}, \vec{r}; x^0, \vec{x}, \vec{x}) \Big]\,,
\end{split}
\end{equation}
where the indices $a$, $b$, $c$, and $d$ are spin indices. This integral contains two terms, each involving a two-particle-state propagator $K_2$ and a one-particle-state propagator $K_1$. The first of them corresponds to the formation of a toponium state at time $x^0$ and position $\vec{x}$, which then propagates as a two-particle system until time $z^0$ and intermediate positions $\vec{r}$ (for the top) and $\vec{z}$ (for the antitop). The antitop then decays in $\vec{z}$ while the top further propagates to $(y^0, \vec{y})$ before decaying. The second term describes instead the reversed sequence where the top decays first. Near threshold and in the presence of QCD interactions, the $t\bar{t}$ system is non-relativistic and bound by the QCD potential $V_\mathrm{QCD}(\vec{r})$, which must thus be included in the computation of $K_2$. In this case, the two-particle propagator is conveniently expressed by separating the centre-of-mass and relative coordinates. Denoting by $x_G$ and $y_G$ as the initial and final centre-of-mass coordinates, and by $x_r$ and $y_r $ the initial and final relative coordinates, 
\begin{equation}
  K_2(y^0, \vec{y}_1, \vec{y}_2; x^0, \vec{x}_1, \vec{x}_2) = \Bigg[ \int \frac{\mathrm{d}^4p}{(2\pi)^4} \, \frac{i}{p^0 - 2m_t - \frac{|\vec{p}|^2}{4 m_t} + i\varepsilon} \, e^{-i p \cdot (y_G - x_G)} \Bigg]\, K_r(y_r; x_r)\,.
\end{equation}
The bracket describes here the free propagation of the toponium centre-of-mass, while $K_r$ is the kernel of the time-dependent Schrödinger equation derived from the non-relativistic QCD Hamiltonian, leading to a Green’s function $G(E; \vec{x})$ in position space defined by~\cite{Strassler:1990nw, Sumino:1992ai}
\begin{equation}
  \left[ -\frac{\vec{\nabla}^2}{m_t} + V_\mathrm{QCD}(\vec{x}) - (E + i \Gamma_t) \right] G(E; \vec{x}) = \delta^{(3)}(\vec{x})\,.
\end{equation}
Here, $E$ is the toponium binding energy and $\Gamma_t$ is the top decay width. This Green's function is in fact the quantity relevant for incorporating toponium effects in predictions, its imaginary part at the origin determining the associated contributions to scattering amplitudes~\cite{Fadin:1987wz, Fadin:1990wx}.

In momentum space, the corresponding Green’s function $\widetilde{G}(E; p^*)$, depending on the binding energy $E$ and the relative momentum $p^*$ of the top and antitop quarks in the $t\bar t$ rest frame, thus plays a central role in incorporating toponium effects into matrix elements describing $t\bar t$ production. It indeed enters through the replacement
\begin{equation}\label{eq:reweighting}
    |\mathcal{M}|^2 \ \to \ |\mathcal{M}|^2 \left|\frac{\widetilde{G}(E; p^*)}{\widetilde{G}_0(E; p^*)}\right|^2\,,
\end{equation}
where $\widetilde{G}_0(E; p^*)$ is the free Green's function and where we consider a process in which a top-antitop pair is produced in the colour-singlet channel. Moreover, we should allow for both on-shell and off-shell top quarks to ensure coverage of the phase space below the $t\bar{t}$ threshold and the correct embedding of spin correlations among final-state particles. In~\cite{Fuks:2021xje}, this re-weighting procedure was applied to matrix elements involving an intermediate pseudo-scalar resonance to model toponium effects, while in~\cite{Maltoni:2024tul}, the same pseudo-scalar toy model was used without re-weighting. However, as will be shown in section~\ref{sec:pheno}, it is more appropriate to re-weight directly matrix elements as calculated in the Standard Model, following the strategy introduced in \cite{Fuks:2024yjj}.

To compute the toponium Green’s function in momentum space, we should solve the Lippmann-Schwinger equation~\cite{Jezabek:1992np, Hagiwara:2016rdv}
\begin{equation}
  \widetilde{G}(E; p) = \widetilde{G}_0(E; p) + \int \frac{\mathrm{d}^3q}{(2\pi)^3} \, \widetilde{V}_\mathrm{QCD}(\vec{p} - \vec{q}) \, \widetilde{G}(E; q)\,,
\end{equation}
which depends on the Fourier transform of the QCD potential $\widetilde{V}_\mathrm{QCD}(\vec{p})$. At short distances and for S-wave toponium production, the potential is dominated by one-gluon exchange and is well approximated by the perturbative Coulomb potential,
\begin{equation}
  \widetilde{V}_\mathrm{Coulomb}(q^2) = -C_F \frac{4\pi\alpha_s(q^2)}{q^2} \left[ 1 + \left(\frac{31}{3} - \frac{10}{9} n_F \right) \frac{\alpha_s(q^2)}{4\pi} \right]\,,
\end{equation}
where $n_F = 5$ is the number of massless quark flavours. In the following, we adopt a tree-level Coulombic potential including a strong coupling constant normalised from its value at the $Z$-pole, and then determine the Green's function $\widetilde G(E; p)$ from $\alpha_s(m_Z) = 0.12$ and a top quark mass $m_t = 173$~GeV and width $\Gamma_t=1.49$~GeV.

\section{Toponium phenomenology at the LHC}\label{sec:pheno}
We now apply the toponium modelling strategy described in section~\ref{sec:topoform} to the case of top-antitop production in LHC collisions at a centre-of-mass energy of $\sqrt{s} = 13$~TeV, followed by a top-antitop decay into a di-leptonic final state,
\begin{equation}\label{eq:process}
    gg \to t\bar{t} \to b \ell^+ \nu_\ell\ \bar{b} \ell^{\prime-} \bar{\nu}_{\ell'}\,.
\end{equation}
As required by the formation of toponium states, we focus on intermediate off-shell $t\bar{t}$ systems of invariant mass $m_{t\bar t} = 2 m_t + E$ and in a colour-singlet configuration, and we additionally restrict our analysis to final states involving electrons and muons for the sake of the example. To simulate this process, we employ the \lstinline{MadGraph5_aMC@NLO} (\lstinline{MG5aMC}) event generator~\cite{Alwall:2014hca}, whose \lstinline{Fortran} output is modified to incorporate the re-weighting from eq.~\eqref{eq:reweighting} and a projection onto a colour-singlet final state, as detailed in~\cite{Fuks:2024yjj}. Event generation is then performed using the modified code, together with the \lstinline{CT18NLO} set of parton distribution functions~\cite{Hou:2019efy, Buckley:2014ana}. Moreover, we include generator-level cuts to restrict our analysis to a phase-space region defined by \mbox{$340$~GeV $\leq m_{t\bar t} \leq 350$~GeV} (corresponding to $-6$~GeV $<E<4$~GeV) and $p^* < 50$~GeV. This ensures that the non-relativistic matrix element, that is required for toponium modelling as the top-antitop pair is non relativistic, is well-approximated by the full matrix element generated automatically by \lstinline{MG5aMC}. Finally, the resulting events are analysed using \lstinline{MadAnalysis 5}~\cite{Conte:2018vmg}. 

\begin{figure}
  \centering
  \includegraphics[width=0.32\linewidth]{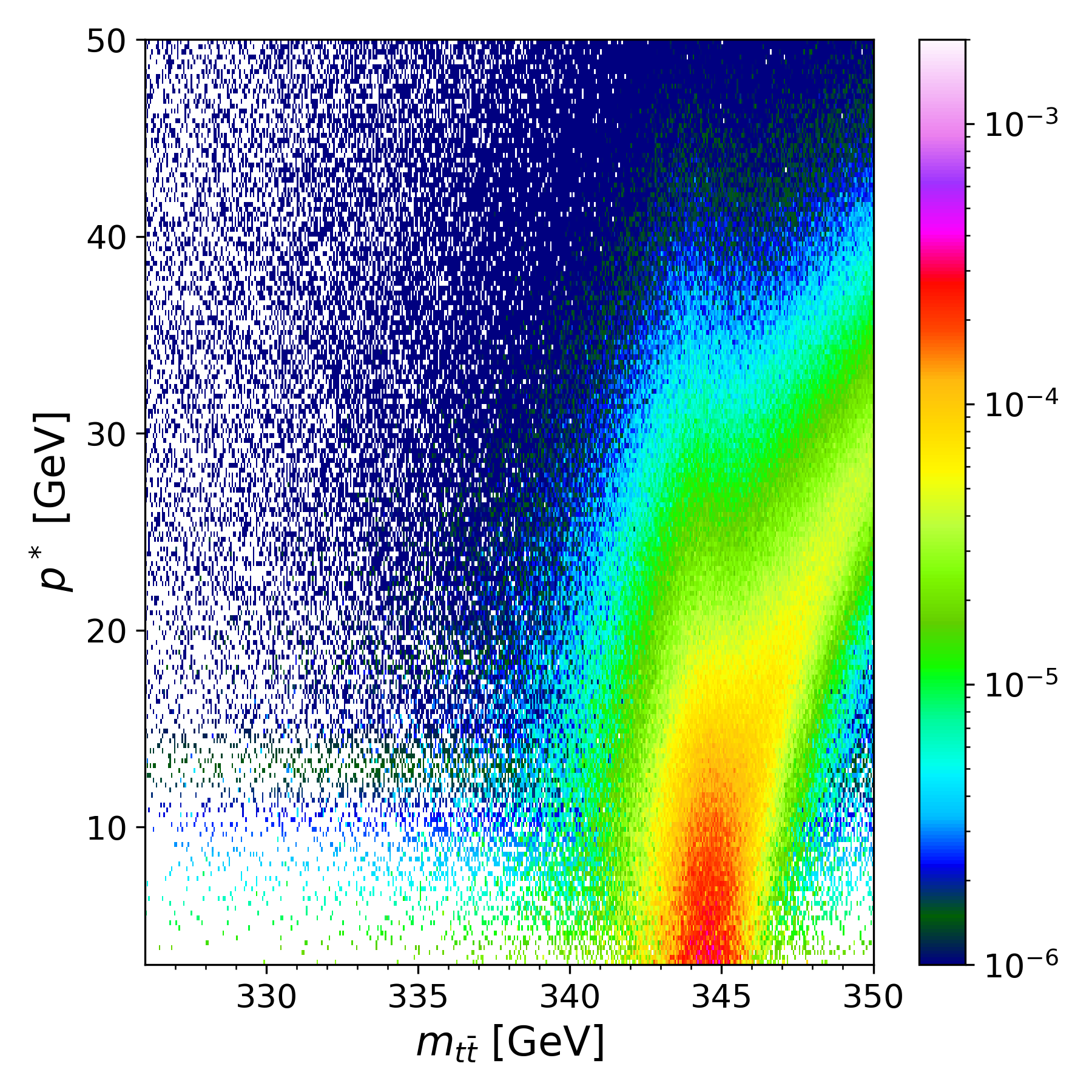}
  \includegraphics[width=0.32\linewidth]{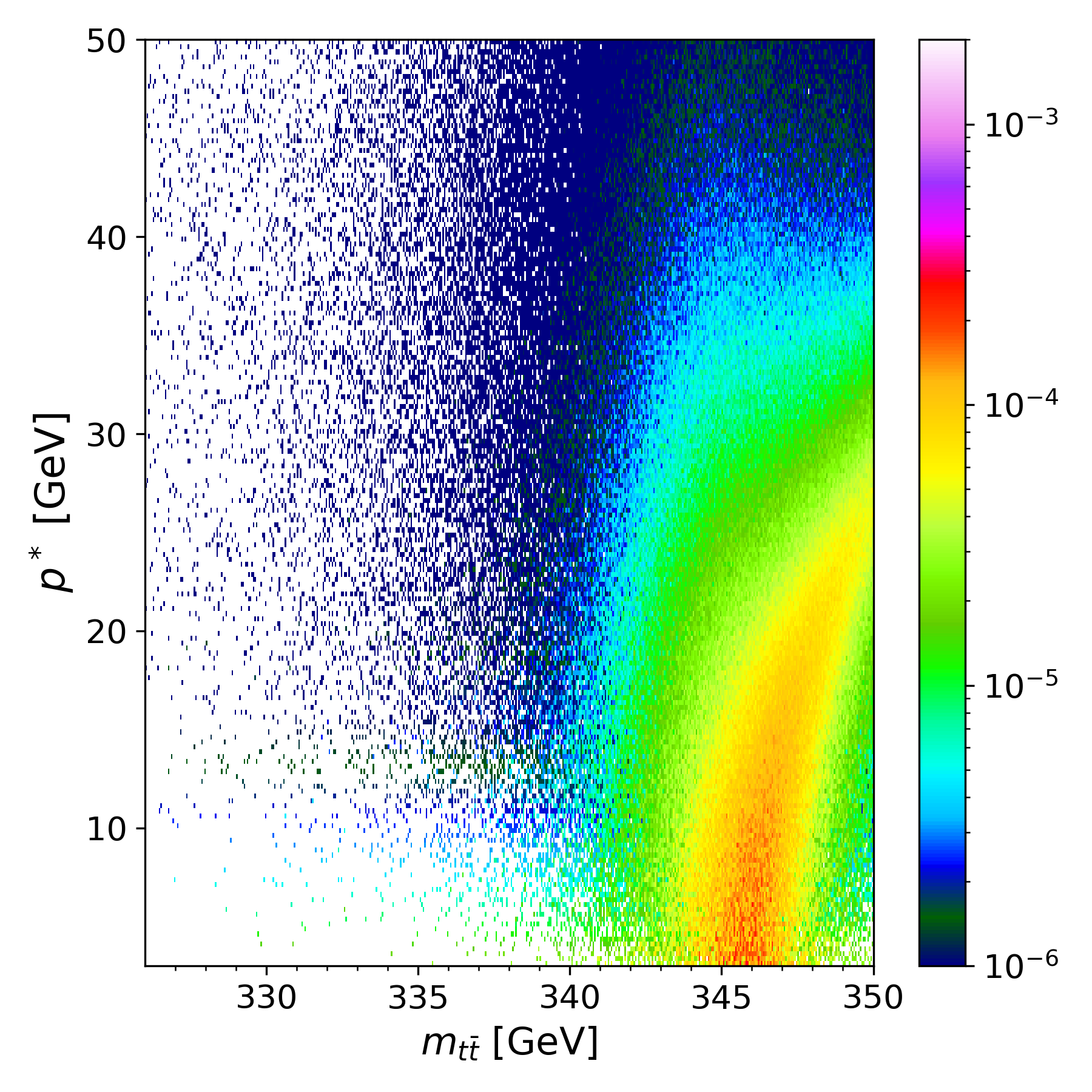}
  \includegraphics[width=0.32\linewidth]{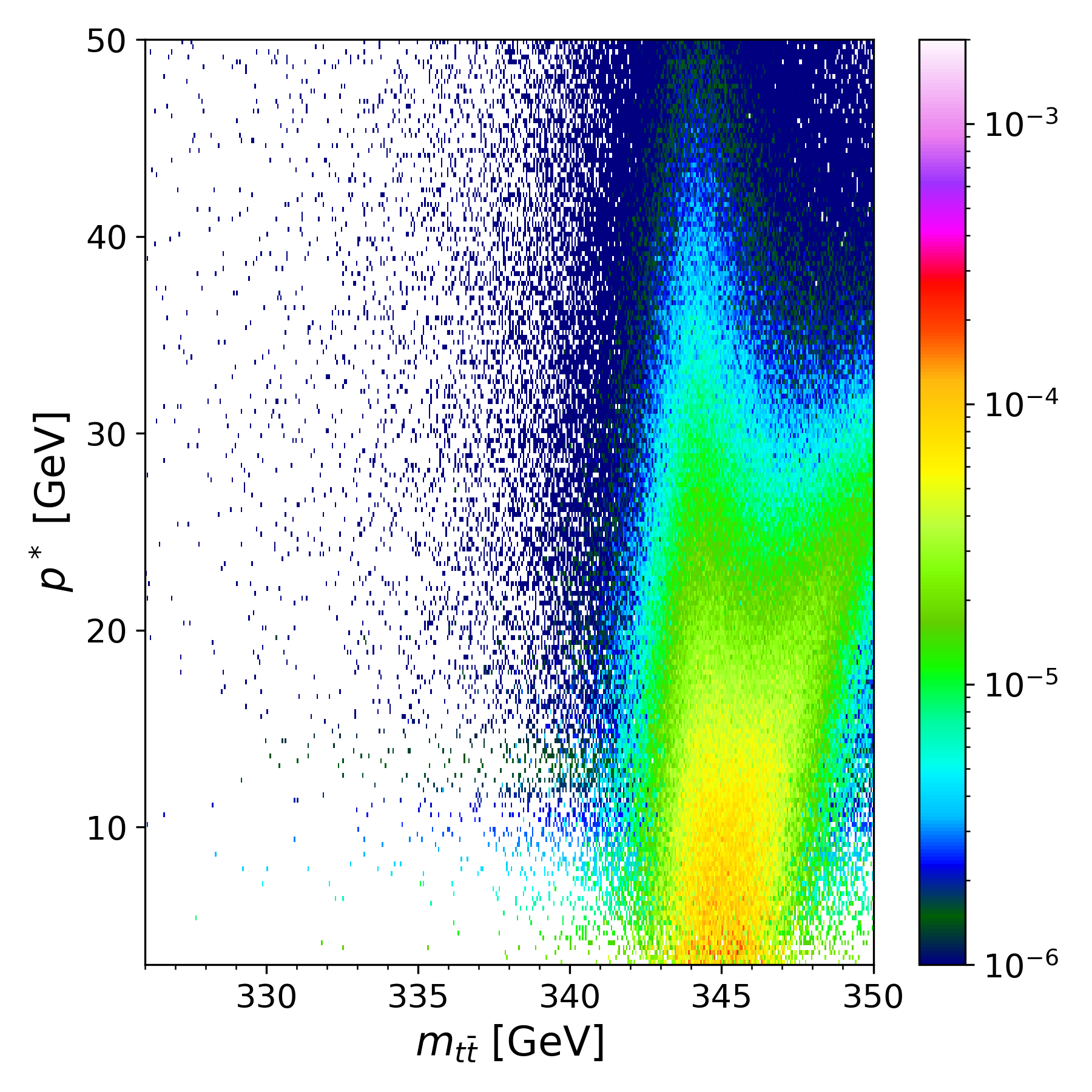}
  \caption{Two-dimensional differential cross section $\mathrm{d}^2\sigma/(p^*)^2$ for colour-singlet top-antitop production at \mbox{$\sqrt{s}=13$~TeV} and di-leptonic decay, as a function of the top quark momentum $p^*$ (in the toponium rest frame) and the invariant mass $m_{t\bar t}$. We compare predictions from non-relativistic QCD (left) with pseudo-scalar toy model predictions with Green's function re-weighting and a toponium width of 7~GeV (centre) and without Green’s function re-weighting and a toponium width of 2.8 GeV (right).\label{fig:d2sigma}}
\end{figure}

The left panel of figure~\ref{fig:d2sigma} presents the doubly-differential cross section in $m_{t\bar t}$ and $p^*$. The distribution's shape follows directly from the properties of the QCD Green’s function, exhibiting a higher density of events around a binding energy of $-2$~GeV and low $p^*$. This behaviour hence reflects the dependence of the magnitude squared of the Green’s function $|\widetilde G(E; p^*)|^2$ on $E$ and $p^*$. Our results thus suggest that precise measurements of this observable at the LHC could give direct access to non-perturbative QCD dynamics. The internal structure of toponium, or in other words the heaviest and most compact hadron in the Standard Model, has thus the potential to become experimentally observable. 

In addition, we also extract from this distribution the average top quark momentum as a function of the binding energy $E$,
\begin{equation}
  \big\langle p(E) \big\rangle  = \frac{\int \mathrm{d}^3p\, p \,\frac{\mathrm{d}^2\sigma}{p^2\, \mathrm{d}p\, \mathrm{d}E}}{\int \mathrm{d}^3p\, \frac{\mathrm{d}^2\sigma}{p^2\, \mathrm{d}p\, \mathrm{d}E}}\,,
\end{equation}
which is approximately 20~GeV at a binding energy $E = -2$~GeV. This is consistent with the non-relativistic QCD expectation~\cite{Sumino:2010bv} and matches the value of the inverse Bohr radius of the top-antitop system,
\begin{equation}
  \frac{1}{a_0} = C_F\, \alpha_s(a_0^{-1})\, \frac{m_t}{2} \,.
\end{equation}
Using $a_0^{-1} \sim 20$~GeV indeed yields a Coulomb potential depth of roughly $-4.5$~GeV, which leads, given a typical kinetic energy of $(p^*)^2/m_t \sim 2.3$~GeV, to a total energy around $-2.2$~GeV consistent with the position of the Green’s function peak.

With the central and right panels of figure~\ref{fig:d2sigma}, we compare the non-relativistic predictions to those obtained from two toy model scenarios proposed in the past, and built from the effective toponium Lagrangian
\begin{equation}
  {\cal L}_{\eta_t} = \frac12 \partial_\mu\eta_t\, \partial^\mu\eta_t
    - \frac12 m_{\eta_t}^2 \eta_t^2
    - \frac{1}{4}g_g\, \eta_t\, G^a_{\mu\nu}\, \tilde G^{a\mu\nu}
    -i g_t\, \eta_t\, \bar{t} \gamma_5 t\,.
\end{equation}
This Lagrangian features a pseudo-scalar toponium field $\eta_t$ of mass $m_{\eta_t}$ coupled to the gluon field strength tensor $G^a_{\mu\nu}$ (and its dual $\tilde G^a_{\mu\nu}$) and the top quark field $t$, with coupling strengths of $g_g$ and $g_t$ respectively. We implement this simplified toy model in \lstinline{MG5aMC} using a UFO module~\cite{Darme:2023jdn} generated with \lstinline{FeynRules}~\cite{Christensen:2009jx, Alloul:2013bka}, and then simulate the process in eq.~\eqref{eq:process} with an intermediate $\eta_t$ resonance using the same toolchain and strategy as described above.

Following~\cite{Fuks:2021xje}, we first take $m_{\eta_t}=344$~GeV and a broad toponium width $\Gamma_{\eta_t}\simeq 7$~GeV, with couplings chosen to match the non-relativistic QCD total cross section of 6.43~pb~\cite{Sumino:2010bv}. This large width is obtained by fitting the difference between the perturbative QCD predictions with and without Coloumbic corrections, and it allows us to capture contributions from both the toponium ground state and excited states. The corresponding predictions for the doubly-differential distribution in $m_{t\bar t}$ and $p^*$ are given in the central panel of figure~\ref{fig:d2sigma}. In contrast, the prescription of~\cite{Maltoni:2024tul} adopts a more physical toponium width, $\Gamma_{\eta_t}=2\Gamma_t \simeq 2.8$~GeV, but it instead omits Green’s function re-weighting. In such a configuration, contributions from higher-spin states are neglected, particularly in the region \mbox{$-0.5$~GeV $\leq E \leq 0$~GeV} where excited S-wave and P-wave states can be found~\cite{Llanes-Estrada:2024phk}. The couplings $g_g$ and $g_t$ are then adjusted to yield once again a total cross section of 6.43~pb. Although both toy models reproduce the toponium peak position near $E \sim -2$~GeV so that the peak position cannot be used as a discriminator in light of current experimental resolution in this phase space region~\cite{CMS:2024ybg, CMS:2025kzt, ATLAS:2023gsl}, they fail to describe the correct momentum spectrum observed in the non-relativistic QCD framework. This highlights the limitations of models based on intermediate resonant propagators with Breit-Wigner shapes, and underscores the relevance of a non-relativistic QCD approach for toponium phenomenology at the LHC. We indeed recall that the $p^*$ distribution controls the structure of the toponium properties as a bound state so that its correct modelling is of utmost importance.

\begin{figure}
  \centering
  \includegraphics[width=0.98\linewidth]{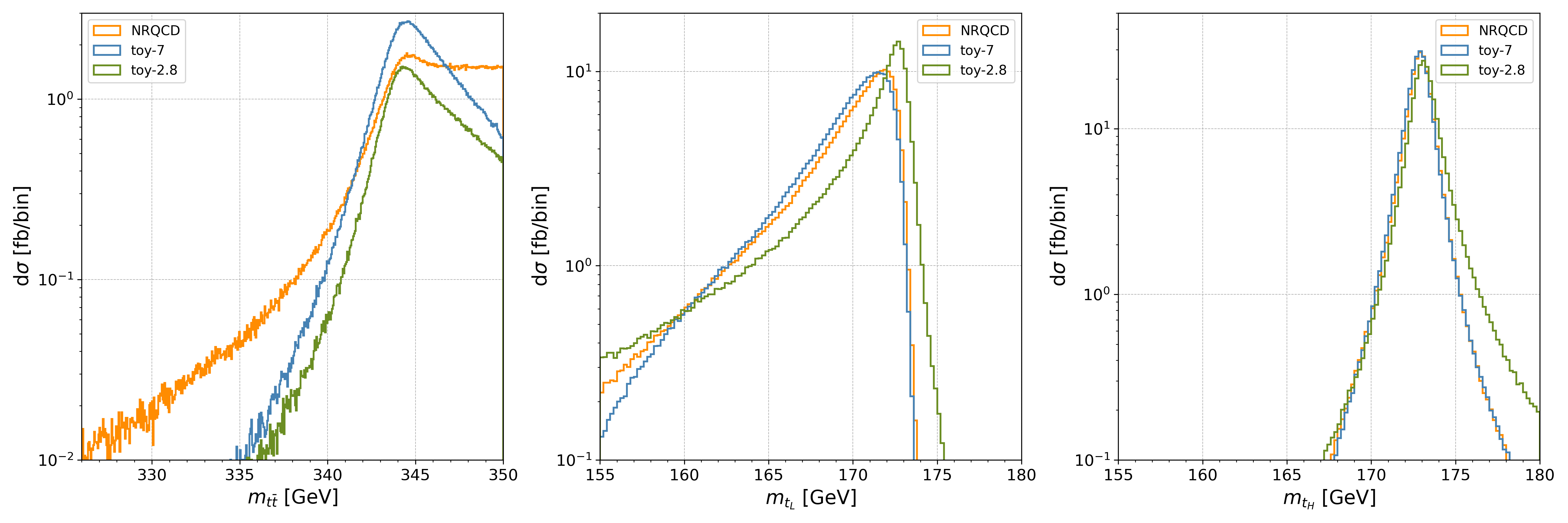}
  \caption{Differential cross section for colour-singlet top-antitop production at \mbox{$\sqrt{s}=13$~TeV} and di-leptonic decay, as a function of the invariant mass of the toponium system (left), the lightest reconstructed top quark (centre), and the heaviest reconstructed top quark (right). Predictions from non-relativistic QCD (orange) are compared to those from pseudo-scalar toy models, both with Green's function re-weighting and a toponium width of 7~GeV (blue), and without re-weighting and a toponium width of 2.8~GeV (green).\label{fig:masses}}
\end{figure}
This is further supported by the results presented in the left panel of figure~\ref{fig:masses}, which shows the invariant mass distribution of the reconstructed top-antitop system obtained by integrating the two-dimensional distribution of figure~\ref{fig:d2sigma} over $p^*$. We compare predictions from non-relativistic QCD (orange) with those from the two toy model configurations: one including the re-weighting of eq.~\eqref{eq:reweighting} and $\Gamma_{\eta_t} = 7$~GeV (blue), and another ignoring any re-weighting and using $\Gamma_{\eta_t} = 2.8$~GeV (green). We observe that all three predictions exhibit a peak consistent with the two-dimensional results and the expectation from the toponium ground states. However, the spectrum shapes differ significantly. In particular, both toy models underestimate the event rate at larger invariant masses, where contributions from higher-spin toponium excitations and octet states should be in order, as shown in~\cite{Kiyo:2008bv,Garzelli:2024uhe}. The discrepancy with the toy models is thus more pronounced in the narrower width scenario, and this should push for the inclusion of both higher-spin and colour-octet contributions in the Green's function.

The role of $p^*$ modelling becomes even clearer in the central and right panels of figure~\ref{fig:masses}, which display the invariant mass distributions of the two reconstructed top quarks, denoted as $t_L$ (lighter) and $t_H$ (heavier). Predictions for the invariant mass of the heavier top quark exhibit a standard Breit-Wigner shape centred at $m_t$ with a width $\Gamma_t$, reflecting the fact that $t_H$ typically decays after $t_L$ and is thus less influenced by bound-state dynamics. In contrast, the invariant mass spectrum of the lighter top quark is distorted and shift to lower values. In the non-relativistic QCD framework, the first quark to decay $t_L$ is treated as a constituent bound by the Coulomb potential of the heavier quark, while $t_H$ remains effectively stable until after $t_L$ has decayed. As a result, $t_L$ exhibits a modified momentum and invariant mass distribution, significantly deviating from the free-particle case. Notably, while toy models reproduce the $m_{t_H}$ distribution fairly well, they fail to accurately capture the dynamics of the lighter quark $t_L$.

\section{Conclusion}
In this report, we examined the impact of toponium formation on top-antitop production at the LHC, comparing various strategies proposed in the literature to model these effects. We showed that previous approaches differ from the non-relativistic QCD expectation, which can be implemented in Monte Carlo event generators by re-weighting the relevant matrix elements using ratios of QCD Green's functions. Our findings support the space-time picture in which a $t\bar{t}$ pair is produced and forms a toponium system with a characteristic size of about $1/20$~GeV$^{-1}$ and a comparable formation time. The system then decays on a timescale of $1/(2\Gamma_t)$, with both the formation and decay occurring well before hadronisation.

Including toponium effects in $t\bar t$ theoretical predictions is therefore essential for improving the precision of the Standard Model predictions, especially in light of recent high-statistics LHC measurements that reveal anomalies potentially linked to the omission of these contributions in simulations. Looking forward, future work should focus on quantifying the associated theoretical uncertainties, extending the formalism to account for higher-spin and higher-partial-wave toponium states and matching predictions to higher-order perturbative calculations.

\section*{Acknowledgements}
This work was supported by the \emph{Agence Nationale de la Recherche} Grant ANR-21-CE31-0013.

\section*{References}
\bibliography{toponium}

\begin{thebibliography}{10}

\bibitem{Fadin:1987wz}
Victor~S. Fadin and Valery~A. Khoze.
\newblock {Threshold Behavior of Heavy Top Production in e+ e- Collisions}.
\newblock {\em JETP Lett.}, 46:525--529, 1987.

\bibitem{ATLAS:2023gsl}
Georges Aad et~al.
\newblock {Inclusive and differential cross-sections for dilepton $
  t\overline{t} $ production measured in $ \sqrt{s} $ = 13 TeV pp collisions
  with the ATLAS detector}.
\newblock {\em JHEP}, 07:141, 2023.

\bibitem{CMS:2024ybg}
Armen Tumasyan et~al.
\newblock {Differential cross section measurements for the production of top
  quark pairs and of additional jets using dilepton events from pp collisions
  at $\sqrt{s}$ = 13 TeV}.
\newblock {\em JHEP}, 02:064, 2025.

\bibitem{CMS:2025kzt}
Aram Hayrapetyan et~al.
\newblock {Observation of a pseudoscalar excess at the top quark pair
  production threshold}.
\newblock {\em CMS-TOP-24-007}, 2025.

\bibitem{Fuks:2021xje}
Benjamin Fuks, Kaoru Hagiwara, Kai Ma, and Ya-Juan Zheng.
\newblock {Signatures of toponium formation in LHC run 2 data}.
\newblock {\em Phys. Rev. D}, 104(3):034023, 2021.

\bibitem{Fadin:1990wx}
Victor~S. Fadin, Valery~A. Khoze, and T.~Sjostrand.
\newblock {On the Threshold Behavior of Heavy Top Production}.
\newblock {\em Z. Phys. C}, 48:613--622, 1990.

\bibitem{Hagiwara:2008df}
Kaoru Hagiwara, Yukinari Sumino, and Hiroshi Yokoya.
\newblock {Bound-state Effects on Top Quark Production at Hadron Colliders}.
\newblock {\em Phys. Lett. B}, 666:71--76, 2008.

\bibitem{Maltoni:2024tul}
Fabio Maltoni, Claudio Severi, Simone Tentori, and Eleni Vryonidou.
\newblock {Quantum detection of new physics in top-quark pair production at the
  LHC}.
\newblock {\em JHEP}, 03:099, 2024.

\bibitem{Fuks:2024yjj}
Benjamin Fuks, Kaoru Hagiwara, Kai Ma, and Ya-Juan Zheng.
\newblock {Simulating toponium formation signals at the LHC}.
\newblock {\em Eur. Phys. J. C}, 85(2):157, 2025.

\bibitem{Sumino:1992ai}
Y.~Sumino, K.~Fujii, Kaoru Hagiwara, H.~Murayama, and C.~K. Ng.
\newblock {Top quark pair production near threshold}.
\newblock {\em Phys. Rev. D}, 47:56--81, 1993.

\bibitem{Strassler:1990nw}
Matthew~J. Strassler and Michael~E. Peskin.
\newblock {The Heavy top quark threshold: QCD and the Higgs}.
\newblock {\em Phys. Rev. D}, 43:1500--1514, 1991.

\bibitem{Jezabek:1992np}
M.~Jezabek, Johann~H. Kuhn, and T.~Teubner.
\newblock {Momentum distributions in t anti-t production and decay near
  threshold}.
\newblock {\em Z. Phys. C}, 56:653--660, 1992.

\bibitem{Hagiwara:2016rdv}
Kaoru Hagiwara, Kai Ma, and Hiroshi Yokoya.
\newblock {Probing CP violation in $e^{+}e^{-}$ production of the Higgs boson
  and toponia}.
\newblock {\em JHEP}, 06:048, 2016.

\bibitem{Alwall:2014hca}
J.~Alwall, R.~Frederix, S.~Frixione, V.~Hirschi, F.~Maltoni, O.~Mattelaer,
  H.~S. Shao, T.~Stelzer, P.~Torrielli, and M.~Zaro.
\newblock {The automated computation of tree-level and next-to-leading order
  differential cross sections, and their matching to parton shower
  simulations}.
\newblock {\em JHEP}, 07:079, 2014.

\bibitem{Hou:2019efy}
Tie-Jiun Hou et~al.
\newblock {New CTEQ global analysis of quantum chromodynamics with
  high-precision data from the LHC}.
\newblock {\em Phys. Rev. D}, 103(1):014013, 2021.

\bibitem{Buckley:2014ana}
Andy Buckley, James Ferrando, Stephen Lloyd, Karl Nordstr\"om, Ben Page, Martin
  R\"ufenacht, Marek Sch\"onherr, and Graeme Watt.
\newblock {LHAPDF6: parton density access in the LHC precision era}.
\newblock {\em Eur. Phys. J. C}, 75:132, 2015.

\bibitem{Conte:2018vmg}
Eric Conte and Benjamin Fuks.
\newblock {Confronting new physics theories to LHC data with MadAnalysis 5}.
\newblock {\em Int. J. Mod. Phys. A}, 33(28):1830027, 2018.

\bibitem{Sumino:2010bv}
Yukinari Sumino and Hiroshi Yokoya.
\newblock {Bound-state effects on kinematical distributions of top quarks at
  hadron colliders}.
\newblock {\em JHEP}, 09:034, 2010.
\newblock [Erratum: JHEP 06, 037 (2016)].

\bibitem{Darme:2023jdn}
Luc Darm\'e et~al.
\newblock {UFO 2.0: the \textquoteleft{}Universal Feynman
  Output\textquoteright{} format}.
\newblock {\em Eur. Phys. J. C}, 83(7):631, 2023.

\bibitem{Christensen:2009jx}
Neil~D. Christensen, Priscila de~Aquino, Celine Degrande, Claude Duhr, Benjamin
  Fuks, Michel Herquet, Fabio Maltoni, and Steffen Schumann.
\newblock {A Comprehensive approach to new physics simulations}.
\newblock {\em Eur. Phys. J. C}, 71:1541, 2011.

\bibitem{Alloul:2013bka}
Adam Alloul, Neil~D. Christensen, C\'eline Degrande, Claude Duhr, and Benjamin
  Fuks.
\newblock {FeynRules 2.0 - A complete toolbox for tree-level phenomenology}.
\newblock {\em Comput. Phys. Commun.}, 185:2250--2300, 2014.

\bibitem{Llanes-Estrada:2024phk}
Felipe~J. Llanes-Estrada.
\newblock {Ensuring that toponium is glued, not nailed}.
\newblock 2024.
\newblock arXiv:2411.19180 [hep-ph].

\bibitem{Kiyo:2008bv}
Y.~Kiyo, Johann~H. Kuhn, S.~Moch, M.~Steinhauser, and P.~Uwer.
\newblock {Top-quark pair production near threshold at LHC}.
\newblock {\em Eur. Phys. J. C}, 60:375--386, 2009.

\bibitem{Garzelli:2024uhe}
M.~V. Garzelli, G.~Limatola, S.~O. Moch, M.~Steinhauser, and O.~Zenaiev.
\newblock {Updated predictions for toponium production at the LHC}.
\newblock 12 2024.
\newblock arXiv:2412.16685 [hep-ph].

\end{thebibliography}

\end{document}